\documentclass[sigconf,screen]{acmart}
\AtBeginDocument{%
  }
    
\setcopyright{cc}
\setcctype{by}
\acmDOI{10.1145/3832783.3834599}
\acmYear{2026}
\copyrightyear{2026}
\acmISBN{979-8-4007-2882-2/2026/10}
\acmConference[ASE '26]{Proceedings of the 41st IEEE/ACM International Conference on Automated Software Engineering}{October 12--16, 2026}{Munich, Germany}
\acmBooktitle{Proceedings of the 41st IEEE/ACM International Conference on Automated Software Engineering (ASE '26), October 12--16, 2026, Munich, Germany}
\acmSubmissionID{ase26tool-p17-p}
\received{2026-05-12}
\received[accepted]{2026-06-19}

\usepackage{microtype}
\usepackage{xcolor}
\usepackage{balance}
\usepackage{url}

\begin{document}

\title{VidTutorAssistant: Automating Responses to Programming Tutorial Questions}
\titlenote{Author's accepted manuscript. Accepted at ASE 2026, Tools and Datasets Track.}

\author{Ahmad Tayeb}
\orcid{0000-0003-4900-729X}
\affiliation{%
  \institution{King Abdulaziz University}
  \city{Jeddah}
  \country{Saudi Arabia}
}
\email{ajtayeb@kau.edu.sa}

\author{Sonia Haiduc}
\correspondingauthor
\orcid{0000-0001-8793-8293}
\affiliation{%
  \institution{Florida State University}
  \city{Tallahassee}
  \country{USA}
}
\email{shaiduc@fsu.edu}

\author{Mohammad D. Alahmadi}
\orcid{0000-0002-3399-2996}
\affiliation{%
  \institution{University of Jeddah}
  \city{Jeddah}
  \country{Saudi Arabia}
}
\email{mdalahmadi@uj.edu.sa}

\renewcommand{\shortauthors}{Ahmad Tayeb, Sonia Haiduc, Mohammad D. Alahmadi}

\begin{abstract}
Programming tutorial videos on YouTube are an important information resource for software developers and students, and their comment sections have evolved into active spaces where viewers ask follow-up questions. The volume of these questions, however, often exceeds what content creators can address, leaving learners without the clarifications they need. We present \textsc{VidTutorAssistant}, a web platform that automates responses to viewer questions on programming video tutorials. \textsc{VidTutorAssistant} implements a retrieval-augmented generation pipeline that extracts a video's transcript, then segments it and embeds it. It then classifies each viewer comment as being a question or non-question, retrieves the most relevant transcript segments to each identified question via cosine similarity, and then generates an answer to the question using an LLM (GPT-4), while grounding the response using the retrieved transcript segments as context. We validate \textsc{VidTutorAssistant} through a study on a subset of 440 user comments selected from a larger dataset of 105{,}553 comments extracted from 7{,}522 Python and Java tutorials. \textsc{VidTutorAssistant} is evaluated on various criteria: a) its ability to identify the programming language in a video, achieving a 0.99 accuracy; b) its ability to classify comments into questions and non-questions, reaching a 0.96 accuracy; and c) its ability to produce correct and complete answers to questions, producing 98\% correct and 99.5\% complete responses, compared with 89\% and 90\% for the original creators' answers. 
\end{abstract}



\maketitle

\section{Introduction}
Programming video tutorials on YouTube have been an important information resource for software developers and computer science students for more than a decade~\cite{macleod2015code, ponzanelli2016codetube}, and they remain a preferred learning resource also in the AI era \cite{tayeb24icsme}. The comment sections of these videos have evolved into active spaces where viewers ask follow-up questions, share feedback, and seek clarifications about specific segments of a video~\cite{poche2017analyzing}. As programming video tutorials accumulate viewers, the volume of questions in their comment sections often exceeds what creators can address, and many learners are left without an answer. This communication gap can slow learning and reduce the educational value of a well-produced tutorial.

Existing work on programming video tutorials focuses primarily on extracting structured artifacts from the videos themselves, including code fragments~\cite{bao2020psc2code}, GUI screens, and workflow actions~\cite{zhao2019actionnet, ponzanelli2016codetube}, rather than on responding to viewer questions. A recent line of research has begun studying question answering for tutorial videos~\cite{yang2024ytcommentqa, yang2024aqua, ray2025eduvidqa}, but these efforts target generic instructional or lecture videos, rely on synthetic or manually anchored questions, and do not release a deployed tool that creators or users of programming tutorials can use directly.

We introduce \textsc{VidTutorAssistant}, a web tool hosted on an online platform that automates responses to viewer questions about programming video tutorials. \textsc{VidTutorAssistant} implements a four-module retrieval-augmented generation (RAG) pipeline. First, a \textit{Video Pre-Processor} module extracts and embeds the video transcript while also identifying the programming language used in the video. Then, a \textit{Comment Classifier} classifies posted user comments into questions and non-questions. A \textit{Context Provider} then reformulates each question into a search query and retrieves the most relevant transcript segments to it via cosine similarity to use as context for the question. A \textit{Response Generator} then prompts GPT-4 with the question and context to produce a contextually grounded reply. One note is that \textsc{VidTutorAssistant} is designed such that GPT-4 can be replaced with any other LLM model, ensuring that the tool remains valuable and relevant as LLMs evolve. The \textsc{VidTutorAssistant} platform is intended both as a standalone tool for viewers and as a draft-generation aid for creators who want to keep their personal voice while reducing the time and cost of replying to viewer questions. \textsc{VidTutorAssistant} is deployed online\footnote{\url{https://vidtutor.softengpaper.com}}, its source code and the dataset we used are available in our replication package\footnote{\url{https://doi.org/10.5281/zenodo.14290285}}, and a screencast of the tool can be found on YouTube\footnote{\url{https://youtu.be/R_OF9XFE1QI}}.

\smallskip\noindent\textbf{Contributions.}
This paper makes the following contributions:
\begin{itemize}\setlength\itemsep{1pt}
  \item A deployed and ready-to-use knowledge-based system called \textsc{VidTutorAssistant} for automated question answering for programming video tutorials, combining RAG with GPT-4.
  \item A dataset of 105{,}553 comments extracted from 7{,}522 Python and Java video tutorials, with a manually labeled evaluation subset of 440 viewer comments and 220 original creator and 220 \textsc{VidTutorAssistant}-associated answers.
  \item A three-part evaluation covering language identification and comment classification accuracy, response correctness and completeness against creator-authored replies, and user-perceived response preference.
\end{itemize}

\section{The VidTutorAssistant Approach}
\label{sec:approach}
\begin{figure}[t]
  \centering

  \includegraphics[width=\linewidth]{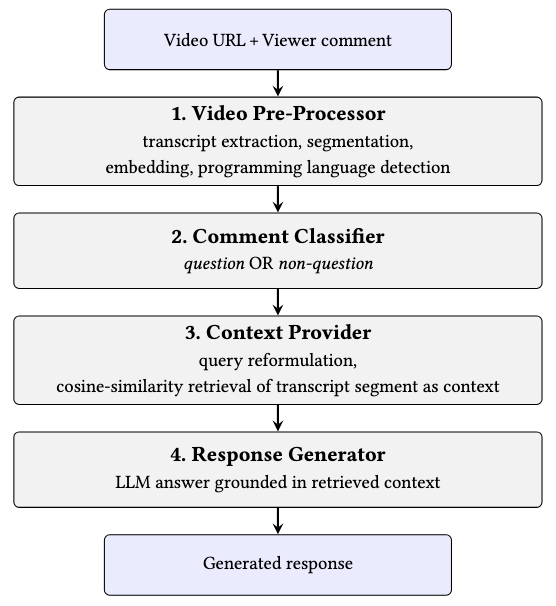}
  \Description{A flowchart connects a video URL and viewer comment to four modules: Video Pre-Processor, Comment Classifier, Context Provider, and Response Generator, ending in a generated response. The caption explains the non-question bypass.}
  \caption{The \textsc{VidTutorAssistant} pipeline. Modules 1--4 run sequentially for \textit{question} comments. \textit{Non-question} comments bypass module~3 and go directly from module~2 to module~4, using the first and last transcript segments as context.}
  \label{fig:approach}
\end{figure}

Figure~\ref{fig:approach} shows the four-module pipeline behind \textsc{VidTutorAssistant}. The \textit{Video Pre-Processor} and \textit{Context Provider} together implement the retrieval side of the RAG pipeline; the \textit{Response Generator} uses the retrieved context to ground the LLM's (GPT-4 in the current implementation) output.

\subsection{Video Pre-Processor}
This module extracts the audio transcript of the input video using the \texttt{youtube-transcript-api} library, identifies the programming language used in the video by prompting the LLM (GPT-4) with the video's title, description, and a transcript excerpt, and prepares the transcript for retrieval. If the transcript exceeds a configurable token threshold (1{,}500 in our deployment), the module splits it into 300-token segments and computes embeddings for each segment using OpenAI's \texttt{text-embedding-3-small} model; otherwise, the full transcript is passed directly to the Response Generator. The 1{,}500-token threshold is chosen in our current deployment in order to leave room for the viewer's comment and the response prompt within GPT-4's context window.

\subsection{Comment Classifier}
This module classifies each viewer comment as being a \textit{question} or \textit{non-question} by prompting the LLM (GPT-4 in our deployment) with a carefully crafted instruction. Comments classified as \textit{non-questions}, such as expressions of thanks or general feedback, bypass the retrieval step, since they do not require a specific video segment to generate a relevant reply. Question comments are passed to the \textit{Context Provider }module for retrieval of relevant context from the video's transcript.

\subsection{Context Provider}
For comments classified as \textit{questions}, this module reformulates the comment into a search query using the LLM (GPT-4), embeds the reformulated query, and computes the cosine similarity of the query to each transcript segment embedding to find the most relevant segments to the query. The video context retrieved and then passed to the \textit{Response Generator} consists of the most relevant segment, its immediate neighbors, and the first and last segments of the video. Including the first and last segments anchors the response in the overall narrative of the video even when the local match alone does not capture the broader context.

\subsection{Response Generator}
This module prompts the LLM (GPT-4) with the viewer's comment, the retrieved video context, and metadata identifying the channel, video, and programming language. The prompt instructs the LLM to formulate an informative response that aligns with the tone and style of the transcript so that responses can serve as drafts that creators may post directly or edit before posting.

\section{Tool Walkthrough}
\label{sec:walkthrough}
\begin{figure}[t]
  \centering
  \includegraphics[width=\linewidth]{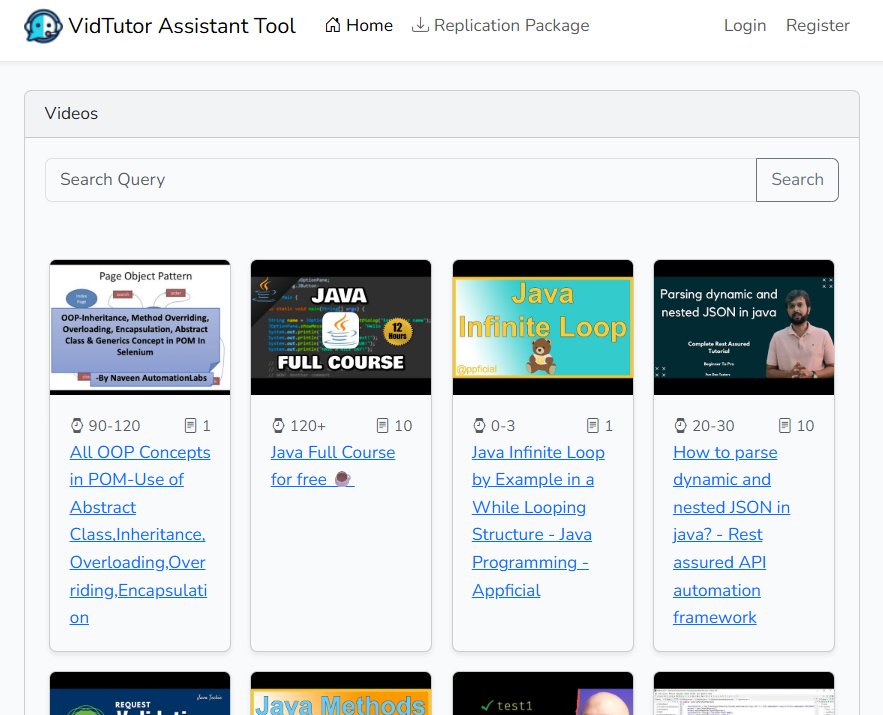}
  \Description{The tutorial catalog has a search field and a grid of video cards showing thumbnails, duration buckets, comment-reply counts, and linked video titles.}
  \caption{\textbf{\textit{Browse Mode.}} The user can search the catalog of programming tutorials using a natural language query. Each video card shows the thumbnail of the video, its duration bucket in minutes, the number of comment-reply pairs for the video, and the video's title, which links to the video's detail page.}
  \label{fig:ui-browse}
\end{figure}

\begin{figure}[t]
  \centering
  \includegraphics[width=\linewidth]{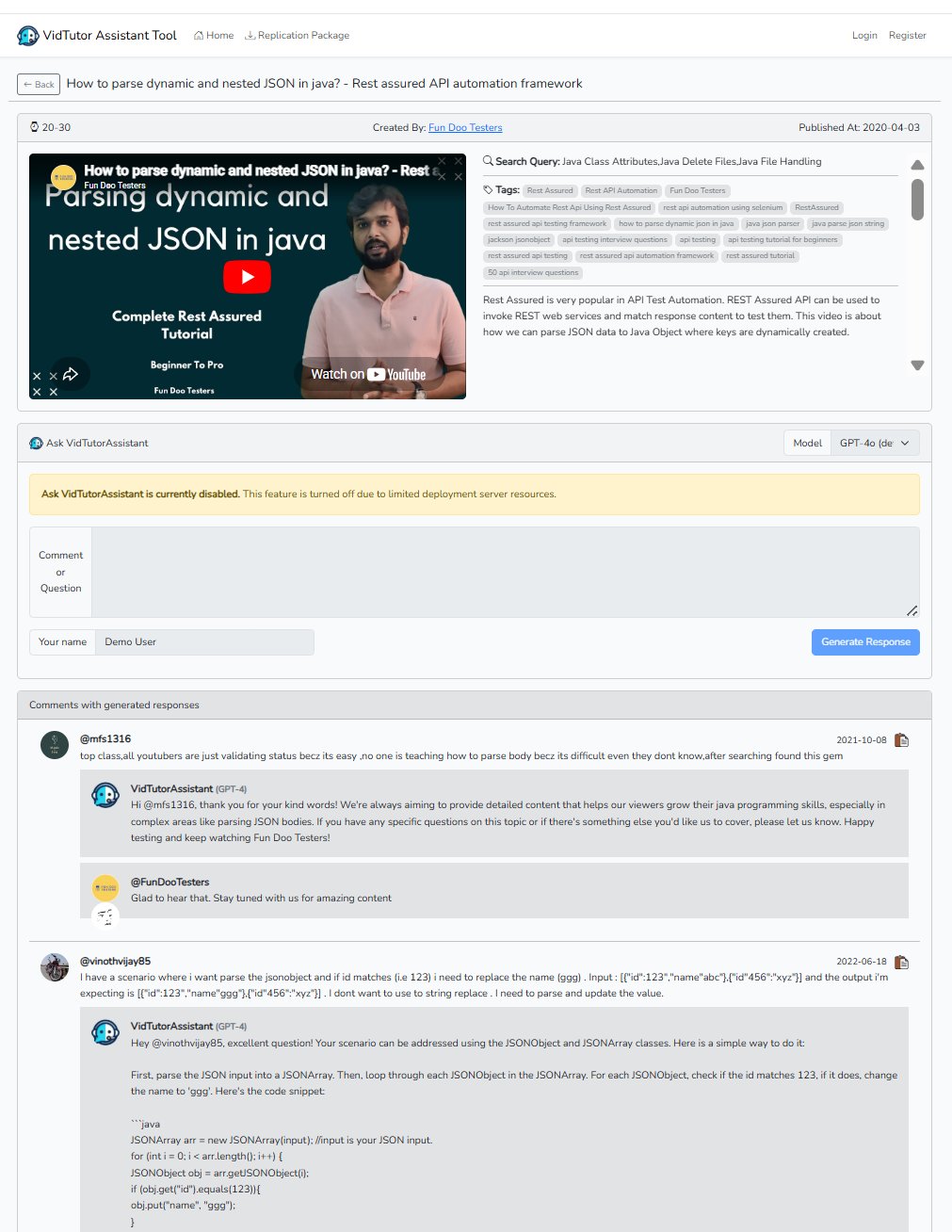}
  \Description{The video detail page shows an embedded tutorial and its metadata, a disabled question widget with a model selector, and stored viewer comments with AI-generated and creator responses.}
  \caption{\textit{Video Detail Mode}. The page shows tutorial metadata and the video, an \emph{Ask VidTutorAssistant} widget for asking new questions, and stored viewer comments paired with their AI-generated responses.}
  \label{fig:ui-detail}
\end{figure}

\textsc{VidTutorAssistant} is deployed at \url{https://vidtutor.softengpaper.com}. The platform has two complementary modes. The \textbf{\emph{Browse Mode}} (Figure~\ref{fig:ui-browse}) lists the catalog of programming tutorials available on the \textsc{VidTutorAssistant} platform. This catalog can be extended as needed. The \textbf{\emph{Video Detail Mode}} (Figure~\ref{fig:ui-detail}) shows, for a selected tutorial, its metadata and embedded video, an \emph{Ask VidTutorAssistant} widget for submitting new questions, and a list of stored viewer comments paired with their generated responses.

\smallskip\noindent\textbf{Browse mode.}
The home page in Figure~\ref{fig:ui-browse} is the platform's entry point. It lists the programming video tutorials available in the dataset as a grid of cards, each containing the video's thumbnail, its duration bucket in minutes (a duration ``bucket'' is a timeframe, e.g., 20--30 minutes), the number of stored comment-reply pairs available for that video, and its title as a clickable link. A free-text search box above the grid allows searching the catalog by title or metadata. The header navigation bar additionally exposes shortcuts to the home page, the replication package, and login and register pages that gate authenticated workflows (e.g., posting feedback or rerunning a query). Selecting a video opens its detail page, described next.

\smallskip\noindent\textbf{Video Detail Mode.}
Selecting a tutorial opens its detail view (Figure~\ref{fig:ui-detail}). The top of the page shows the video's title, channel, publication date, duration bucket, search queries used to discover the video during dataset construction, YouTube tags, and description, alongside the embedded original video. An \emph{Ask VidTutorAssistant} widget below the metadata lets a viewer submit a free-text comment or question and pick the LLM that should generate the response (e.g., GPT-4, GPT-4o); in our current public deployment, this widget is disabled to limit server load, while the offline pipeline continues to populate stored responses. The lower portion of the page lists the stored viewer comments for the video, each paired with the \textsc{VidTutorAssistant}-generated response. Where the original video creator also replied to a comment on YouTube, that reply is displayed beneath the AI response, mirroring the side-by-side comparison setup of the user study reported in Section~\ref{sec:validation}.

\smallskip\noindent\textbf{Implementation.}
The platform follows a two-tier architecture. A Python backend runs pipeline modules 1--4 (transcript extraction, classification, context retrieval, LLM answer generation), while the user-facing interface is a Laravel (PHP) web application with a Blade and Livewire frontend that forwards user inputs to the Python API and renders the returned responses. The \emph{Ask VidTutorAssistant} widget supports multiple LLM models through a model selector; currently available models (i.e., GPT-4 and GPT-4o) are accessed through the OpenAI API. Transcript embeddings are produced with \texttt{text-embedding-3-small} and cached per video. Generated responses, evaluation metadata, and user feedback are stored in MySQL on the web.

\section{Evaluation}
\label{sec:validation}
We validate \textsc{VidTutorAssistant} with three studies. \textit{Study~1} evaluates the auxiliary modules 1 and 2, specifically the programming-language identification and comment classification capabilities. \textit{Study~2} evaluates the correctness and completeness of the responses generated by \textsc{VidTutorAssistant} and compares them to the original creator-authored replies extracted from YouTube in our dataset. \textit{Study~3} is a user study comparing participant preferences between the two types of responses (original creator vs  \textsc{VidTutorAssistant}).

\smallskip\noindent\textbf{Dataset.}
Table~\ref{tab:dataset} summarizes our dataset, available on the   \textsc{VidTutorAssistant} platform and in the replication package. We collected comment-reply pairs from Python and Java tutorials on YouTube using a list of 55 Java and 78 Python topics extracted from W3Schools\footnote{https://www.w3schools.com/} as seed queries. After retaining only English videos with transcripts and pairing each top-level comment with the first reply provided by the original creator, we obtained 105,553 comment-reply pairs from 7,522 unique videos. We then sampled a representative evaluation subset of 440 comments, balanced across the two programming languages, eleven different video-duration buckets, and the \textit{question} vs.\ \textit{non-question} categories of comments, with at most five comments drawn from any single video. Two software developers, holding MS degrees in computer science, independently labeled each comment as being a \textit{question} or \textit{non-question}, with near-perfect agreement (Cohen's $\kappa = 0.97$).

\begin{table}[t]
  \centering
  \caption{Summary of the \textsc{VidTutorAssistant} dataset.}
  \label{tab:dataset}
  \small
  \setlength{\tabcolsep}{4pt}
  \begin{tabular}{@{}ll@{}}
    \toprule
    \textbf{Property} & \textbf{Value} \\
    \midrule
    Source videos                          & 7,522 \\
    Programming languages                  & 2 (Python, Java) \\
    Comment-reply pairs                    & 105,553 \\
    Manually labeled evaluation subset     & 440 \\
    \quad Question / non-question split    & 220 / 220 \\
    Evaluated answer pairs (creator + tool)& 220 \\
    User study participants                & 46 \\
    User study reviews                     & 438 \\
    \bottomrule
  \end{tabular}
\end{table}

\smallskip\noindent\textbf{Methodology.}
\textit{Study~1} evaluates the \textit{Video Pre-Processor}'s programming language predictions and the \textit{Comment Classifier}'s question / non-question classifications against the manually labeled ground truth on the 440-comment subset extracted from a total of 82 videos. \textit{Study~2} focuses the evaluation on the 220 question comments and asks two programmers holding MS degrees in computer science to independently rate each pair of answers (one from the original creator and one from \textsc{VidTutorAssistant}) on correctness and completeness; to avoid bias, the source of each answer (creator or \textsc{VidTutorAssistant}) was hidden during the evaluation. A third evaluator resolved disagreements, leading to a final Cohen's $\kappa = 0.86$. \textit{Study~3} is a within-subject user study with 46 computer science students, recruited via a departmental mailing list. Each participant expressed their preference between the \textsc{VidTutorAssistant} and creator responses on a subset of 173 questions for which both answers had been previously confirmed correct and complete by our evaluators; participants also provided a free-text rationale for their evaluations, which we analyzed via open coding (Krippendorff's $\alpha = 0.93$).

\smallskip\noindent\textbf{Results.}
Programming language identification reached 0.99 accuracy (precision 1.00, recall 0.99, F1 0.99); 81 of 82 evaluated videos are correctly identified, with one classified as ``unknown.'' Comment classification achieved 0.96 accuracy with precision, recall, and F1 all at 0.96; 422 of 440 comments are correctly classified.

On the 220 evaluated question comments, \textsc{VidTutorAssistant} produces 98\% correct (216/220), 1\% partially correct, and 1\% incorrect responses, compared to 89\% correct (195/220), 8\% partially correct, and 4\% incorrect for the original creators' responses. \textsc{VidTutorAssistant}'s responses are also rated 99.5\% complete (219/220) compared to only 90\% (198/220) of the creators' answers.

\begin{figure}[t]
  \centering
  \includegraphics[width=\linewidth]{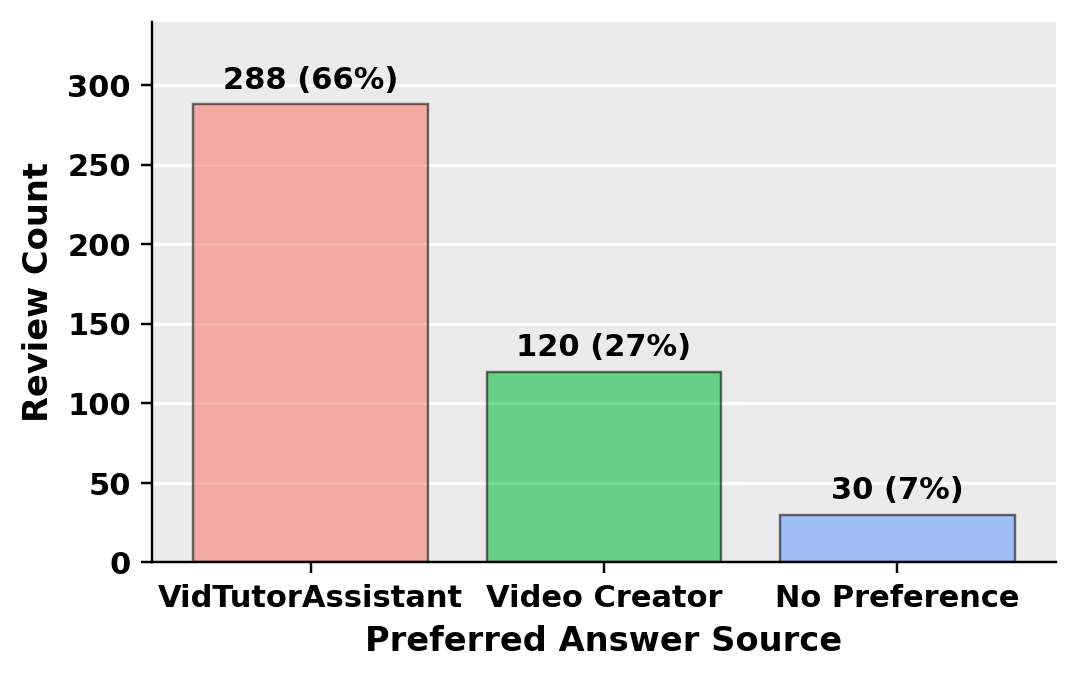}
  \Description{A bar chart shows 288 reviews preferring VidTutorAssistant, 120 preferring the video creator, and 30 expressing no preference, corresponding to 66, 27, and 7 percent of 438 reviews.}
  \caption{Participant preferences between \textsc{VidTutorAssistant} and the original creator responses across 438 reviews by 46 computer science students.}
  \label{fig:preference}
\end{figure}

In the user study, participants preferred \textsc{VidTutorAssistant}'s responses 66\% of the time (288 of 438 reviews), the creators' responses 27\% of the time (120 of 438), and reported no preference for the remaining 7\% (30 of 438), as shown in Figure~\ref{fig:preference}. The 66\% versus 27\% split is significant against the null of equal preference under a one-sample binomial test ($p < 0.001$ over the 408 reviews that expressed a preference). Open coding of the free-text rationales revealed that the users' preference for \textsc{VidTutorAssistant}'s answers is primarily due to the detail and comprehensiveness of its answers, clearer explanations, and a friendlier tone, while creators were preferred when their conciseness suited a simple question.

Overall, the studies indicate that \textsc{VidTutorAssistant}'s retrieval-augmented pipeline produces responses that are at least as correct and complete as those of the original creators and that students prefer the \textsc{VidTutorAssistant}'s answers more often than those of the original creators.

\section{Related Tools}
The most related prior tools to \textsc{VidTutorAssistant} fall into two main categories. First are tools that extract structured artifacts from tutorial videos, such as CodeTube~\cite{ponzanelli2016codetube}, ActionNet~\cite{zhao2019actionnet}, and psc2code~\cite{bao2020psc2code}. However, they do not address the comments and questions posted by viewers. The second category is that of LLM-based question-answering tools for tutorial videos~\cite{yang2024ytcommentqa,yang2024aqua,ray2025eduvidqa}; however, YTCommentQA~\cite{yang2024ytcommentqa} classifies answerability rather than generating answers, AQuA~\cite{yang2024aqua} targets non-programming feature-rich software via manual visual anchors, and EduVidQA~\cite{ray2025eduvidqa} benchmarks multimodal LLMs on synthetic and curated lecture-video QA pairs. \textsc{VidTutorAssistant} differs by focusing on programming tutorials, evaluating against the original creator replies, and being released as a deployed platform.

\section{Conclusion and Future Work}
We presented \textsc{VidTutorAssistant}, a deployed web platform that automates responses to viewer questions on programming video tutorials using a retrieval-augmented GPT-4 pipeline; on 220 evaluated questions it produces 98\% correct and 99.5\% complete responses, and computer science students prefer its replies 66\% of the time. Planned extensions cover broader language support beyond Python and Java, creator-side draft review workflows, and comparison against newer LLMs (including open-source and multimodal models).

\section{Data Availability Statement}
\textsc{VidTutorAssistant} is freely available and deployed online (\url{https://vidtutor.softengpaper.com}), its source code and the dataset we used are available in our replication package (\url{https://doi.org/10.5281/zenodo.14290285}), and a screencast of the tool can be found on YouTube (\url{https://youtu.be/R_OF9XFE1QI}).

\begin{acks}
This work was supported in part by the US National Science Foundation grant 1846142. 
\end{acks}

\balance

\bibliographystyle{ACM-Reference-Format}
\bibliography{references}

\end{document}